\PassOptionsToPackage{unicode}{hyperref}
\PassOptionsToPackage{hyphens}{url}
\PassOptionsToPackage{dvipsnames,svgnames,x11names}{xcolor}
\documentclass[
]{article}

\usepackage{amsmath,amssymb}
\usepackage{iftex}
\ifPDFTeX
  \usepackage[T1]{fontenc}
  \usepackage[utf8]{inputenc}
  \usepackage{textcomp} % provide euro and other symbols
\else % if luatex or xetex
  \usepackage{unicode-math}
  \defaultfontfeatures{Scale=MatchLowercase}
  \defaultfontfeatures[\rmfamily]{Ligatures=TeX,Scale=1}
\fi
\usepackage{lmodern}
\ifPDFTeX\else  
\fi
\IfFileExists{upquote.sty}{\usepackage{upquote}}{}
\IfFileExists{microtype.sty}{% use microtype if available
  \usepackage[]{microtype}
  \UseMicrotypeSet[protrusion]{basicmath} % disable protrusion for tt fonts
}{}
\makeatletter
\@ifundefined{KOMAClassName}{% if non-KOMA class
  \IfFileExists{parskip.sty}{%
    \usepackage{parskip}
  }{% else
    \setlength{\parindent}{0pt}
    \setlength{\parskip}{6pt plus 2pt minus 1pt}}
}{% if KOMA class
  \KOMAoptions{parskip=half}}
\makeatother
\usepackage{xcolor}
\ifx\paragraph\undefined\else
  \let\oldparagraph\paragraph
  \renewcommand{\paragraph}[1]{\oldparagraph{#1}\mbox{}}
\fi
\ifx\subparagraph\undefined\else
  \let\oldsubparagraph\subparagraph
  \renewcommand{\subparagraph}[1]{\oldsubparagraph{#1}\mbox{}}
\fi

\providecommand{\tightlist}{%
  \setlength{\itemsep}{0pt}\setlength{\parskip}{0pt}}\usepackage{longtable,booktabs,array}
\usepackage{calc} % for calculating minipage widths
\usepackage{etoolbox}
\makeatletter
\patchcmd\longtable{\par}{\if@noskipsec\mbox{}\fi\par}{}{}
\makeatother
\IfFileExists{footnotehyper.sty}{\usepackage{footnotehyper}}{\usepackage{footnote}}
\makesavenoteenv{longtable}
\usepackage{graphicx}
\makeatletter
\def\maxwidth{\ifdim\Gin@nat@width>\linewidth\linewidth\else\Gin@nat@width\fi}
\def\maxheight{\ifdim\Gin@nat@height>\textheight\textheight\else\Gin@nat@height\fi}
\makeatother
\setkeys{Gin}{width=\maxwidth,height=\maxheight,keepaspectratio}
\makeatletter
\def\fps@figure{htbp}
\makeatother
\NewDocumentCommand\citeproctext{}{}

\makeatletter
 \let\@cite@ofmt\@firstofone
 \def\@biblabel#1{}
 \def\@cite#1#2{{#1\if@tempswa , #2\fi}}
\makeatother
\newlength{\cslhangindent}
\newlength{\csllabelwidth}
\newenvironment{CSLReferences}[2] % #1 hanging-indent, #2 entry-spacing
 {\begin{list}{}{%
  \setlength{\itemindent}{0pt}
  \setlength{\leftmargin}{0pt}
  \setlength{\parsep}{0pt}
  \ifodd #1
   \setlength{\leftmargin}{\cslhangindent}
   \setlength{\itemindent}{-1\cslhangindent}
  \fi
  \setlength{\itemsep}{#2\baselineskip}}}
 {\end{list}}
\usepackage{calc}

\usepackage{arxiv}
\usepackage{orcidlink}
\usepackage{amsmath}
\usepackage[T1]{fontenc}
\makeatletter
\@ifpackageloaded{caption}{}{\usepackage{caption}}
\AtBeginDocument{%
\ifdefined\contentsname
  \renewcommand*\contentsname{Table of contents}
\else
  \newcommand\contentsname{Table of contents}
\fi
\ifdefined\listfigurename
  \renewcommand*\listfigurename{List of Figures}
\else
  \newcommand\listfigurename{List of Figures}
\fi
\ifdefined\listtablename
  \renewcommand*\listtablename{List of Tables}
\else
  \newcommand\listtablename{List of Tables}
\fi
\ifdefined\figurename
  \renewcommand*\figurename{Figure}
\else
  \newcommand\figurename{Figure}
\fi
\ifdefined\tablename
  \renewcommand*\tablename{Table}
\else
  \newcommand\tablename{Table}
\fi
}
\@ifpackageloaded{float}{}{\usepackage{float}}
\floatstyle{ruled}
\@ifundefined{c@chapter}{\newfloat{codelisting}{h}{lop}}{\newfloat{codelisting}{h}{lop}[chapter]}
\floatname{codelisting}{Listing}

\makeatother
\makeatletter
\@ifpackageloaded{caption}{}{\usepackage{caption}}
\@ifpackageloaded{subcaption}{}{\usepackage{subcaption}}
\makeatother
\ifLuaTeX
  \usepackage{selnolig}  % disable illegal ligatures
\fi
\usepackage{bookmark}

\IfFileExists{xurl.sty}{\usepackage{xurl}}{} % add URL line breaks if available
\hypersetup{
  pdftitle={scores: A Python package for verifying and evaluating models and predictions with xarray},
  colorlinks=true,
  linkcolor={blue},
  filecolor={Maroon},
  citecolor={Blue},
  urlcolor={Blue},
  pdfcreator={LaTeX via pandoc}}

\renewcommand{\today}{July 3, 2024}
\newcommand{\runninghead}{A Preprint }
\renewcommand{\runninghead}{A Preprint - scores: A Python package for
verifying and evaluating models and predictions with xarray }
\title{scores: A Python package for verifying and evaluating models and
predictions with xarray}
\def\asep{\\\\\\ } % default: all authors on same column
\def\asep{\And }
\author{\textbf{Tennessee
Leeuwenburg}~\orcidlink{0009-0008-2024-1967}\\\\Bureau of Meteorology,
Australia\\\\\href{mailto:tennessee.leeuwenburg@bom.gov.au}{tennessee.leeuwenburg@bom.gov.au}\asep\textbf{Nicholas
Loveday}~\orcidlink{0009-0000-5796-7069}\\\\Bureau of Meteorology,
Australia\\\\\asep\textbf{Elizabeth E. Ebert}\\\\Bureau of Meteorology,
Australia\\\\\asep\textbf{Harrison
Cook}~\orcidlink{0009-0009-3207-4876}\\\\Bureau of Meteorology,
Australia\\\\\asep\textbf{Mohammadreza
Khanarmuei}~\orcidlink{0000-0002-5017-9622}\\\\Bureau of Meteorology,
Australia\\\\\asep\textbf{Robert J.
Taggart}~\orcidlink{0000-0002-0067-5687}\\\\Bureau of Meteorology,
Australia\\\\\asep\textbf{Nikeeth
Ramanathan}~\orcidlink{0009-0002-7406-7438}\\\\Bureau of Meteorology,
Australia\\\\\asep\textbf{Maree
Carroll}~\orcidlink{0009-0008-6830-8251}\\\\Bureau of Meteorology,
Australia\\\\\asep\textbf{Stephanie
Chong}~\orcidlink{0009-0007-0796-4127}\\\\Independent Contributor,
Australia\\\\\asep\textbf{Aidan Griffiths}\\\\Work undertaken while at
the Bureau of Meteorology, Australia\\\\\asep\textbf{John
Sharples}\\\\Bureau of Meteorology, Australia\\\\}
\date{July 3, 2024}
\begin{document}
\maketitle

\section{Summary}\label{summary}

\texttt{scores} is a Python package containing mathematical functions
for the verification, evaluation and optimisation of forecasts,
predictions or models. It supports labelled n-dimensional
(multidimensional) data, which is used in many scientific fields and in
machine learning. At present, \texttt{scores} primarily supports the
geoscience communities; in particular, the meteorological,
climatological and oceanographic communities.

\texttt{scores} not only includes common scores (e.g., Mean Absolute
Error), it also includes novel scores not commonly found elsewhere
(e.g., FIxed Risk Multicategorical (FIRM) score, Flip-Flop Index),
complex scores (e.g., threshold-weighted continuous ranked probability
score), and statistical tests (such as the Diebold Mariano test). It
also contains isotonic regression which is becoming an increasingly
important tool in forecast verification and can be used to generate
stable reliability diagrams. Additionally, it provides pre-processing
tools for preparing data for scores in a variety of formats including
cumulative distribution functions (CDF). At the time of writing,
\texttt{scores} includes over 50 metrics, statistical techniques and
data processing tools.

All of the scores and statistical techniques in this package have
undergone a thorough scientific and software review. Every score has a
companion Jupyter Notebook tutorial that demonstrates its use in
practice.

\texttt{scores} supports \texttt{xarray} datatypes, allowing it to work
with Earth system data in a range of formats including NetCDF4, HDF5,
Zarr and GRIB among others. \texttt{scores} uses Dask for scaling and
performance. Support for \texttt{pandas} is being introduced.

The \texttt{scores} software repository can be found at
\url{https://github.com/nci/scores/}.

\pagebreak

\section{Statement of Need}\label{statement-of-need}

Labelled, n-dimensional data is widely used in many scientific fields.
The Earth system science community makes heavy use of physics-based and
machine learning models, both to process observations (such as
identifying land use from satellite data) and to make predictions about
the future (such as forecasting the weather). These models, predictions
and forecasts undergo verification and evaluation to assess their
correctness.

The purpose of \texttt{scores} is (a) to mathematically verify and
validate models and predictions and (b) to foster research into new
scores and metrics.

\texttt{scores} handles dimensionality and weighting (e.g., latitude
weighting) more effectively than commonly-used data science packages.
While there are existing open source Python verification packages for
labelled n-dimensional data (see ``Related Software Packages'' further
below), none of these packages offer all of the key benefits of
\texttt{scores}.

\subsection{\texorpdfstring{Key Benefits of
\texttt{scores}}{Key Benefits of scores}}\label{key-benefits-of-scores}

To meet the needs of researchers and other users, \texttt{scores}
provides the following key benefits.

\subsubsection{Data Handling}\label{data-handling}

\begin{itemize}
\tightlist
\item
  Works with labelled, n-dimensional data (e.g., geospatial, vertical
  and temporal dimensions) for both point-based and gridded data.
  \texttt{scores} can effectively handle the dimensionality, data size
  and data structures commonly used for:

  \begin{itemize}
  \tightlist
  \item
    gridded Earth system data (e.g., numerical weather prediction
    models)
  \item
    tabular, point, latitude/longitude or site-based data (e.g.,
    forecasts for specific locations).
  \end{itemize}
\item
  Handles missing data, masking of data and weighting of results.
\item
  Supports \texttt{xarray} (Hoyer \& Hamman, 2017) datatypes, and works
  with NetCDF4 (Unidata, 2024), HDF5 (The HDF Group \& Koziol, 2020),
  Zarr (Miles et al., 2020) and GRIB (World Meteorological Organization,
  2024) data formats among others.
\end{itemize}

\subsubsection{Usability}\label{usability}

\begin{itemize}
\tightlist
\item
  A companion Jupyter Notebook (Jupyter Team, 2024) tutorial for each
  metric and statistical test that demonstrates its use in practice.
\item
  Novel scores not commonly found elsewhere (e.g., FIRM (Taggart et al.,
  2022), Flip-Flop Index (Griffiths et al., 2019, 2021)).
\item
  Commonly-used scores are also included, meeting user requests to use
  \texttt{scores} as a standalone package.
\item
  All scores and statistical techniques have undergone a thorough
  scientific and software review.
\item
  An area specifically to hold emerging scores which are still
  undergoing research and development. This provides a clear mechanism
  for people to share, access and collaborate on new scores, and be able
  to easily re-use versioned implementations of those scores.
\end{itemize}

\subsubsection{Compatibility}\label{compatibility}

\begin{itemize}
\tightlist
\item
  Highly modular - provides its own implementations, avoids extensive
  dependencies and offers a consistent API.
\item
  Easy to integrate and use in a wide variety of environments. It has
  been used on workstations, servers and in high performance computing
  (supercomputing) environments.
\item
  Maintains 100\% automated test coverage.
\item
  Uses Dask (Dask Development Team, 2016) for scaling and performance.
\item
  Support for \texttt{pandas} (McKinney, 2010; The pandas development
  team, 2024) is being introduced.
\end{itemize}

\pagebreak

\subsection{\texorpdfstring{Metrics, Statistical Techniques and Data
Processing Tools Included in
\texttt{scores}}{Metrics, Statistical Techniques and Data Processing Tools Included in scores}}\label{metrics-statistical-techniques-and-data-processing-tools-included-in-scores}

At the time of writing, \texttt{scores} includes over 50 metrics,
statistical techniques and data processing tools. For an up to date
list, please see the \texttt{scores} documentation.

The ongoing development roadmap includes the addition of more metrics,
tools, and statistical tests.

\hfill \break

\begin{longtable}[]{@{}
  >{\raggedright\arraybackslash}p{(\columnwidth - 4\tabcolsep) * \real{0.1667}}
  >{\raggedright\arraybackslash}p{(\columnwidth - 4\tabcolsep) * \real{0.2024}}
  >{\raggedright\arraybackslash}p{(\columnwidth - 4\tabcolsep) * \real{0.6310}}@{}}
\caption{A \textbf{curated selection} of the metrics, tools and
statistical tests currently included in \texttt{scores}}\tabularnewline
\toprule\noalign{}
\begin{minipage}[b]{\linewidth}\raggedright
\end{minipage} & \begin{minipage}[b]{\linewidth}\raggedright
\textbf{Description}
\end{minipage} & \begin{minipage}[b]{\linewidth}\raggedright
\textbf{A Selection of the Functions Included in \texttt{scores}}
\end{minipage} \\
\midrule\noalign{}
\endfirsthead
\toprule\noalign{}
\begin{minipage}[b]{\linewidth}\raggedright
\end{minipage} & \begin{minipage}[b]{\linewidth}\raggedright
\textbf{Description}
\end{minipage} & \begin{minipage}[b]{\linewidth}\raggedright
\textbf{A Selection of the Functions Included in \texttt{scores}}
\end{minipage} \\
\midrule\noalign{}
\endhead
\bottomrule\noalign{}
\endlastfoot
\textbf{Continuous} & Scores for evaluating single-valued continuous
forecasts. & Mean Absolute Error (MAE), Mean Squared Error (MSE), Root
Mean Squared Error (RMSE), Additive Bias, Multiplicative Bias, Pearson's
Correlation Coefficient, Flip-Flop Index (Griffiths et al., 2019, 2021),
Quantile Loss, Murphy Score (Ehm et al., 2016). \\
& & \\
\textbf{Probability} & Scores for evaluating forecasts that are
expressed as predictive distributions, ensembles, and probabilities of
binary events. & Brier Score (Brier, 1950), Continuous Ranked
Probability Score (CRPS) for Cumulative Distribution Functions (CDFs)
(including threshold-weighting, see Gneiting \& Ranjan (2011)), CRPS for
ensembles (Ferro, 2013; Gneiting \& Raftery, 2007), Receiver Operating
Characteristic (ROC), Isotonic Regression (reliability diagrams)
(Dimitriadis et al., 2021). \\
& & \\
\textbf{Categorical} & Scores (including contingency table metrics) for
evaluating forecasts of categories. & Probability of Detection (POD),
Probability of False Detection (POFD), False Alarm Ratio (FAR), Success
Ratio, Accuracy, Peirce's Skill Score (Peirce, 1884), Critical Success
Index (CSI), Gilbert Skill Score (Gilbert, 1884), Heidke Skill Score,
Odds Ratio, Odds Ratio Skill Score, F1 Score, Symmetric Extremal
Dependence Index (Ferro \& Stephenson, 2011), FIxed Risk
Multicategorical (FIRM) Score (Taggart et al., 2022). \\
& & \\
\textbf{Spatial} & Scores that take into account spatial structure. &
Fractions Skill Score (Roberts \& Lean, 2008). \\
& & \\
\textbf{Statistical Tests} & Tools to conduct statistical tests and
generate confidence intervals. & Diebold-Mariano (Diebold \& Mariano,
1995) with both the Harvey et al. (1997) and Hering \& Genton (2011)
modifications. \\
& & \\
\textbf{Processing Tools} & Tools to pre-process data. & Data matching,
discretisation, cumulative density function manipulation. \\
\end{longtable}

\subsection{Use in Academic Work}\label{use-in-academic-work}

In 2015, the Australian Bureau of Meteorology began developing a new
verification system called Jive, which became operational in 2022. For a
description of Jive see Loveday, Griffiths, et al. (2024). The Jive
verification metrics have been used to support several publications
(Foley \& Loveday, 2020; Griffiths et al., 2017; Taggart, 2022a, 2022b,
2022c).

\texttt{scores} has arisen from the Jive verification system and
provides Jive verification functions as a modular, open source package.
\texttt{scores} also includes additional metrics that Jive does not
contain.

\texttt{scores} has been used to explore user-focused approaches to
evaluating probabilistic and categorical forecasts (Loveday, Taggart, et
al., 2024).

\subsection{Related Software Packages}\label{related-software-packages}

There are multiple open source verification packages in a range of
languages. Below is a comparison of \texttt{scores} to other open source
Python verification packages. None of these include all of the metrics
implemented in \texttt{scores} (and vice versa).

\texttt{xskillscore} (Bell et al., 2021) provides many but not all of
the same functions as \texttt{scores}. The Jupyter Notebook tutorials in
\texttt{scores} cover a wider array of metrics.

\texttt{climpred} (Brady \& Spring, 2021) uses \texttt{xskillscore}
combined with data handling functionality, and is focused on ensemble
forecasts for climate and weather. \texttt{climpred} makes some design
choices related to data structure (specifically associated with climate
modelling) which may not generalise effectively to broader use cases.
Releasing \texttt{scores} separately allows the differing design
philosophies to be considered by the community.

\texttt{METplus} (Brown et al., 2021) is a substantial verification
system used by weather and climate model developers. \texttt{METplus}
includes a database and a visualisation system, with Python and shell
script wrappers to use the \texttt{MET} package for the calculation of
scores. \texttt{MET} is implemented in C++ rather than Python.
\texttt{METplus} is used as a system rather than providing a modular
Python API.

\texttt{Verif} (Nipen et al., 2023) is a command line tool for
generating verification plots whereas \texttt{scores} provides a Python
API for generating numerical scores.

\texttt{Pysteps} (Imhoff et al., 2023; Pulkkinen et al., 2019) is a
package for producing short-term ensemble predictions, focusing on
probabilistic nowcasting of radar precipitation fields. It includes a
significant verification submodule with many useful verification scores.
\texttt{Pysteps} does not provide a standalone verification API.

\texttt{PyForecastTools} (Morley \& Burrell, 2020) is a Python package
for model and forecast verification which supports \texttt{dmarray}
rather than \texttt{xarray} data structures and does not include Jupyter
Notebook tutorials.

\section{Acknowledgements}\label{acknowledgements}

We would like to thank Jason West and Robert Johnson from the Bureau of
Meteorology for their feedback on an earlier version of this manuscript.

We would like to thank and acknowledge the National Computational
Infrastructure (nci.org.au) for hosting the \texttt{scores} repository
within their GitHub organisation.

\section*{References}\label{references}
\addcontentsline{toc}{section}{References}

\phantomsection\label{refs}

\end{document}